\documentclass[twocolumn,prl,aps,showpacs]{revtex4}
\usepackage{graphicx}
\usepackage{dcolumn}
\usepackage{bm}

\begin{document}

\title{Size dependent tunneling and optical spectroscopy of CdSe quantum rods}

\author{David Katz}
\author{Tommer Wizansky}
\author{Oded Millo}
\email{milode@vms.huji.ac.il} \affiliation{Racah Institute for
Physics, and the Center for Nanoscience and Nanotechnology, The
Hebrew University of Jerusalem, Jerusalem 91904, Israel}
\author{Eli Rothenberg}
\author{Taleb Mokari}
\author{Uri Banin}
\email{banin@chem.ch.huji.ac.il} \affiliation{Institute of
Chemistry, and the Center for Nanoscience and Nanotechnology, The
Hebrew University of Jerusalem, Jerusalem 91904, Israel}

\begin{abstract}
  Photoluminescence excitation spectroscopy and scanning tunneling
  spectroscopy are used to study the electronic states in CdSe
  quantum rods that manifest a transition from a zero dimensional
  to a one dimensional quantum confined structure.  Both optical
  and tunneling spectra show that the level structure depends
  primarily on the rod diameter and not on length. With
  increasing diameter, the band-gap and the excited state level
  spacings shift to the red.  The level structure was assigned
  using a multi-band effective-mass model, showing a similar
  dependence on rod dimensions.
\end{abstract}
\maketitle

Colloidal semiconductor nanocrystals are a class of nanomaterials
that manifest the transition from the molecular limit to the solid
state \cite{Alivisatos,MilloPRL}, with significant potential for
serving as building blocks of nano-devices in applications ranging
from lasers \cite{KlimovScienceEtal,BaninAdvMatEtal} and
opto-electronic devices \cite{UriNirScience} to biological
fluorescence tagging \cite{AlivisatosBioEtal}. Shape control of
such colloidally prepared nanostructures has been recently
achieved by modifying the synthesis to obtain rod shaped particles
- quantum rods \cite{PengNatureEtal}. Quantum rods (QRs) exhibit
electronic and optical properties different than QDs.  For
example, unlike the spherical dots, QRs have linearly polarized
emission as demonstrated recently by fluorescence measurements on
single rods \cite{HuScienceEtal}, leading to polarized lasing
\cite{BaninAdvMatEtal}. In this letter we combine optical and
tunneling spectroscopies to investigate the electronic level
structure of CdSe quantum rods, and study its dependence on rod
length and diameter. The levels are assigned with the use of a
multi-band effective-mass model. The study provides significant
insight on the evolution of the electronic structure from zero
dimensional QDs to one-dimensional quantum wires.

The combination of scanning tunneling spectroscopy with PLE has
proven to be a powerful approach to decipher the level structure
of nanocrystal QDs \cite{MilloPRL,BaninNature}.  While in the
optical spectra, allowed valence band (VB) to conduction band (CB)
transitions are detected \cite{BawendiPRB,BaninPhysChemEtal}, in
tunneling spectroscopy the CB and VB states can be separately
probed yielding complimentary information on the level structure
\cite{BaninNature, BakkersRapid, VanmaekelberghNanoLettEtal,
KatzAPL}. These data provide an important benchmark for testing
theoretical models developed for describing the level structure in
strongly quantum-confined nanostructures. This was demonstrated
for spherical nanocrystal QDs where multi-band effective mass
approaches \cite{BawendiPRB, BaninPhysChemEtal, EfrosJOSABEtal}
and atomistic pseudo-potential theory \cite{Zunger} were both used
to describe measured size-dependent level structure. In a recent
work by Hu et al. \cite{HuPseudoEtal}, the pseudo-potential
approach was also applied for the calculation of energy levels in
rods of small diameter and aspect ratios up to six.

The CdSe system studied here is highly developed synthetically, to
a level where it serves as a model system for colloidal
semiconductor nanostructures.  Fig. 1 presents  transmission
electron microscopy (TEM) images of typical samples of CdSe
quantum rods grown using the well developed methods of colloidal
nanocrystal synthesis
\cite{PengNatureEtal,AlivisatosJACS,PengJACS}.
  For the study of length and diameter dependence we prepared six
different samples with dimensions that can be divided into two
groups- three samples with radii $\sim1.8$ nm and lengths ranging
from 11 to 31 nm, and three samples, with radii $\sim3.2$ nm and
lengths ranging from 11 to 60 nm.  For the low temperature optical
experiments, optically clear free-standing polymer films
containing the rods were prepared. The cooled samples showed
absorption spectra with several transitions and had a distinct
photoluminescence (PL) peak assigned to band-gap emission.  The
positions of the absorption onset and PL peak red shifted with
increasing diameter and showed no significant variation with rod
length [Fig. 4(a)] \cite{AlivisatosNanoLett}.

\begin{figure}
\resizebox{6cm}{!} {\includegraphics{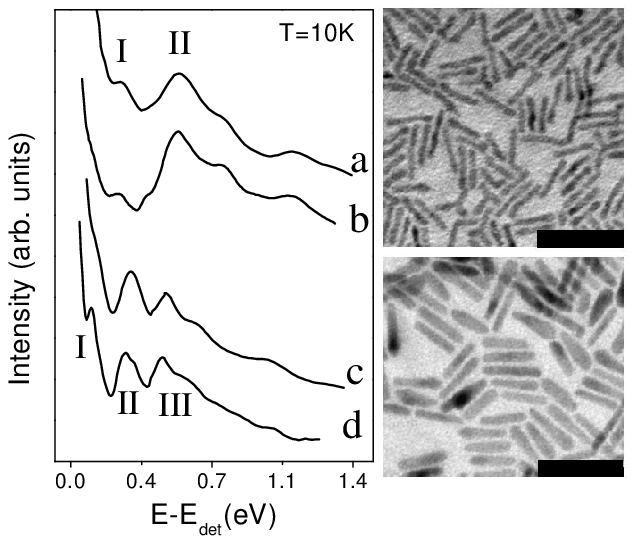}} \caption{PLE
spectra for CdSe QRs are shown in the left frame (a) 31x1.9
(length times radius) nm, $E_{det}=2.25$ eV (b) 11x1.6 nm,
$E_{det}=2.25$ eV (c) 60x3.3 nm, $E_{det}=2.00$ eV (d) 11x2.9 nm,
$E_{det}=2.03$ eV, with the zero energy representing the position
of the detection window, $E_{det}$. Relevant optical transitions
are marked.  The structure above 0.7 eV is overlapping peaks of
the excitation lamp that could not be completely normalized out.
TEM images of two QR samples, $31\times 3.9$ nm (top), and $30
\times 3.2$ nm (bottom), are shown in the rigth. Scale bar is 50
nm. } \label{fig1}
\end{figure}

The PLE spectra presented in Fig. 1, measured on four samples,
were obtained by opening a narrow detection window at the blue
edge of the inhomogeneously broadened PL peak
\cite{BawendiPRB,BaninPhysChemEtal}.
 The spectra are less structured as compared with those
measured on QDs due to the increased sources of inhomogeneous
broadening in rods and due to the intrinsically less discrete
level structure. A striking feature is the nearly identical level
structure observed for the two QR samples of small diameter
(traces a,b), which differs significantly from the spectra
measured on the thicker rods (traces c,d), although in each group
the aspect ratios vary from 3 to about 10. This shows clearly that
not only the band-gap of the rods depends mainly on the diameter,
but also the excited optical transitions.

The optical measurements were accompanied by tunneling
measurements performed at $4.2$ $K$ on $single$ QRs deposited on
graphite, using a cryogenic scanning tunneling microscope (STM)
\cite{KatzAPL}.  A topographic image of a nanorod 25 nm in length
and 2 nm in radius is presented in Fig. 2(a).  After identifying
such an isolated QR, tunneling $dI/dV$ vs. $V$ spectra were
obtained in a double barrier tunnel junction (DBTJ) configuration
\cite{BaninNature, KatzAPL, BakkersRapid,
VanmaekelberghNanoLettEtal}, as shown schematically in Fig. 2(b).
The spectra were measured with the tip retracted from the QR to a
distance where the bias voltage is dropped largely across the
tip-QR junction, and charging effects are avoided. As a result, CB
(VB) states appear at positive (negative) sample bias, and the
peak separations are close to the real QR level spacings. Hence,
the $dI/dV$ spectra yield direct information on the QR level
spectrum \cite{KatzAPL, BakkersRapid, VanmaekelberghNanoLettEtal}.

The tunneling spectra in Fig. 2(c) further demonstrate, in
accordance with the PLE data, that the QR level structure depends
primarily on the diameter of the QRs, not on their length. Most
significantly, the region of suppressed tunneling conductance
(null density of states) around zero bias, associated with the
quasi-particle energy gap, is red shifted upon QR thickening, from
$\sim2.4$ eV in the upper two curves ($r \sim2$ nm) to $\sim2.2$
eV in the lower curves ($r\sim2.7$ mn). This trend was not
observed as clearly for the spacing between the CB ground state
(CB1) and the first excited state (CB2), as evident also from Fig.
4(b). Level CB3 appearing in the lower trace was observed in
$\sim50$\% of the measurements, where the current did not reach
the saturation limit before resolving this level.  The VB is
considerably more dense and complex (see below) and its level
structure was not reliably resolved in our tunneling spectra; we
thus only denote the first (ground state) peak as VB1 [Fig. 2(c)].
More insight into the QR level structure, including the excited VB
levels, is gained by correlating the tunneling spectra with
allowed optical VB to CB transitions and comparing these data with
model level-structure calculations.

To calculate the electronic structure of the rods we employ the
formalism developed by Sercel and Vahala for quantum wires
\cite{Vahala}.  Similar basis functions were used for the envelope
wavefunctions, with quantum numbers reflecting the cylindrical
symmetry of the QR: $n$, the principal number (associated with the
number of nodes along the radius), and $F_{z}$ and $L_{z}$, the
projections of the total and the envelope angular momenta along
the z (symmetry) axis, respectively.  Here, $F_{z}$ = $J_{z}$ +
$L_{z}$, where $J_{z}$  is the projection of band-edge Bloch
angular momentum along the z axis. The finite length $L$ of the
QRs is expressed by introducing an additional quantum number,
denoted by $m$, and a related factor, $exp(i\pi mz/L)$, to the
envelope wavefunctions, replacing the continuum wavevector $K_{z}$
in Ref. \onlinecite{Vahala}.

We first consider the CB level structure. Here, due to the large
band-gap of CdSe (1.84 eV), the uncoupled one-band effective-mass
model with quantum numbers ($n,L_{z},m$) is adequate for rods of
large enough aspect ratio \cite{Vahala, HuPseudoEtal}. Figure 3(a)
presents CB energy levels calculated using this model for a
particle in a finite cylindrical potential-well ($U = 5$ eV), as a
function of radius and length (inset).  The CB states with $L_{z}
= 0,1,$ and 2 are denoted CB1, CB2 and CB3, respectively. This
figure clearly depicts the strong dependence of the energy levels
on the QR radius, due to the strong confinement associated with
this 'narrow dimension', as opposed to the near lack of length
dependence shown in the upper inset. The length dependence becomes
significant only for small aspect ratio (around 2), or for high
$m$ values, effectively shortening the QR, since the number of
nodes along the z axis is $m-1$, in agreement with Ref.
\onlinecite{HuPseudoEtal}.

\begin{figure}
\resizebox{6cm}{!} {\includegraphics{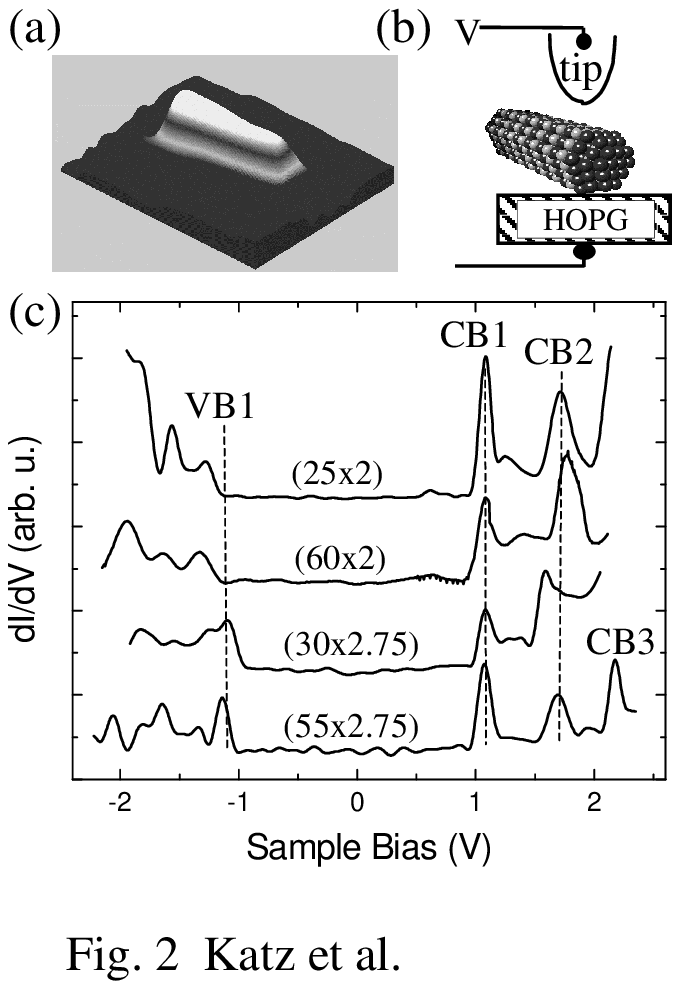}} \caption{(a) STM
topographic image of a QR 25 nm long and 2 nm in radius. (b)
Schematic of the DBTJ configuration used for acquiring the
tunneling spectra. (c) $dI/dV - V$ tunneling spectra for QRs of
various lengths and radii, marked above each curve in nm. For
clarity, the spectra were shifted horizontally to align the CB1
peaks.  The vertical dashed lines are guides to the eye. The
tunneling set values were in the range $V_{s}=1.5-2$ eV and
$I_{s}=50-80$ pA.} \label{fig2}
\end{figure}

The VB states were calculated using the four-band effective-mass
model that couples heavy and light holes \cite{Vahala}, where the
quantum numbers are now ($n,F_{z},m$). Results of this
calculation, performed for an infinite cylindrical potential-well,
are presented in Fig. 3(b), where states with $F_{z} = 1/2, 3/2,$
and 5/2 (and $m=1$) are shown. Again, the near lack of dependence
of the level structure on the rod length is manifested by the
inset.  It is evident that the VB is significantly more dense as
compared to the CB. Yet, the VB levels near the band-edge can be
divided into three groups: VB1, denoting the two upper levels,
VB2, the three subsequent ones and VB3, the next group.  Each of
these groups is composed of various $F_{z}$ levels, where, in
turn, each $F_{z}$ level has contributions from both heavy and
light holes ($J_{z}=±3/2$ and ±1/2, respectively), and thus from
wavefunctions of different $L_{z}$ values. This VB mixing leads to
very weak selection rules for the optical transitions.

In Fig. 4 we compare the measured optical transitions and
tunneling spectra with our theoretical calculations. The gap
extracted from the optical measurements (solid circles) and the
gap identified in the tunneling data (empty circles), are plotted
in Fig. 4(a) along with the calculated energy gap, VB1-CB1.  In
order to compare with the quasi-particle (tunneling) gap, the
measured excitonic (optical) gap was corrected for the
electron-hole Coulomb interaction. As a first approximation, we
modified the expression given in Ref. \onlinecite{BawendiPRB} for
QDs, $1.8e^{2}/kr$ (where $k$ is the dielectric constant), to take
into account the fact that in our QRs strong confinement holds
only for the radial dimension. We thus replace the radius, $r$, by
$(r^{2}a_{o})^{1/3}$, where $a_{o}$ is the Bohr radius, 5.7 nm.  A
relatively good agreement between the calculated and measured
energy gaps is found, for both tunneling and optical experiments.
However, the tunneling gaps are higher in energy and show a
stronger dependence on radius as compared to the optical data, as
was previously observed for QDs \cite{BaninNature, BawendiPRB,
BaninPhysChemEtal}. This can be ascribed to the non-vanishing
voltage drop on the QR-substrate junction \cite{KatzAPL,
BakkersRapid, VanmaekelberghNanoLettEtal} and on the uncertainty
involved in the determination of the QR radii \cite{BaninNature}.
Concerning the optical gap, the transition from both $F_{z}=1/2$
and 3/2 VB1 states are allowed. In the absorption they are too
close to be resolved, but we note that the state with $F_{z}=1/2$
is the top-most state, from which theory predicts a transition
polarized along the symmetry axis \cite{SercelWirePolarization},
consistent with PL polarization measurements for QRs
\cite{HuScienceEtal}.

Turning now to the excited levels, Fig. 4(b) shows spacings of the
PLE transitions I and II with respect to the band-gap transition,
together with levels CB2 and CB3 measured with respect to CB1,
detected by tunneling.  The data are presented as a function of
the band-gap measured in each experiment, thus eliminating the
possible problem of QR radius estimation mentioned above. Most
interesting is the good agreement of the spacing between PLE
transition II and the band-gap transition (solid squares) with the
tunneling data for the CB2-CB1 level separation (open squares),
allowing a clear assignment of optical transition II. Both the
optical and tunneling data correlate with the theoretical curve
for the CB2-CB1 spacing.  It is also clear now that PLE transition
I (solid stars) can only take place between an excited VB state
and the CB ground state.  However, due to the intricate VB level
structure and the broad PLE features, it cannot be unambiguously
assigned, and relatively good agreement is found for both
calculated VB2-VB1 and VB3-VB1 level spacing (dashed lines).
Transition III was resolved only for the three samples with larger
diameters, and thus cannot be assigned. Considering again the
tunneling data, Fig. 4(b) shows a weaker size (energy-gap)
dependence for CB2 and CB3 as compared to theory.  This may be
partly due to our oversimplified model, where we assume a sharp
(square-well) and somewhat large (5 eV) confining potential. These
approximations should affect mainly the higher levels.

\begin{figure}
\resizebox{6cm}{!} {\includegraphics{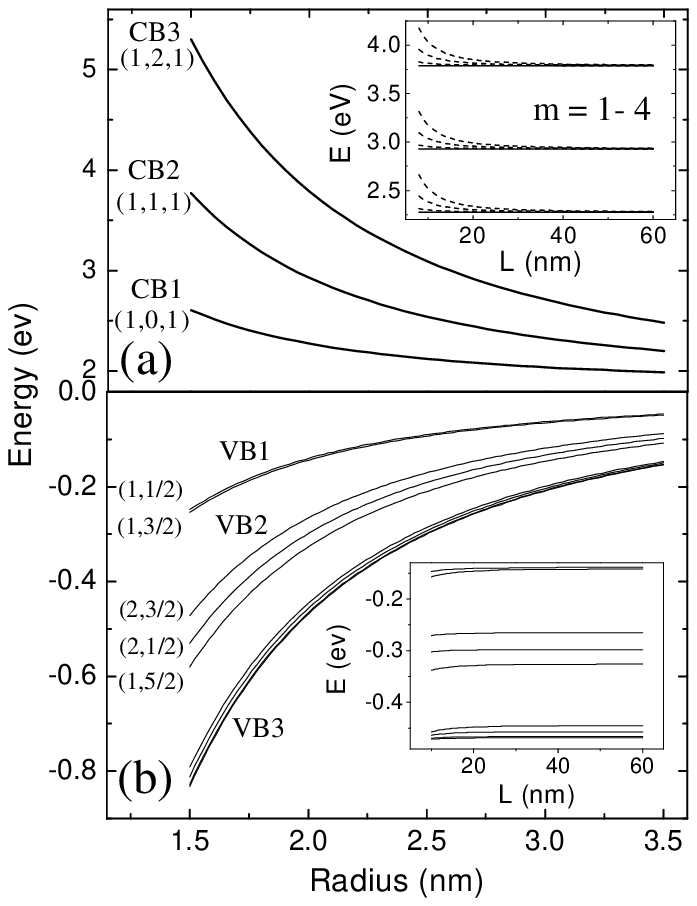}}
\caption{Calculated energy levels of CdSe QRs. (a) First three CB
states versus nanorod radius, for a 30 nm long QR. The quantum
numbers $(n,L_{z},m)$ are denoted. (b) The VB energy levels versus
QR radius calculated for a 30 nm long rod. The quantum numbers
$(n,F_{z})$ are denoted $(m=1)$. Group VB3 consists of states with
$F_{z}$ = 1/2, 3/2, 5/2, and 7/2. Insets: Length dependence for
the respective energy levels for a rod 2 nm in radius. The upper
inset shows also the dependence on the $m$ quantum number. The
energies are given with respect to the bulk VB edge. Effective
masses for electrons and holes and the bulk energy gap were taken
from Ref. \onlinecite{EfrosJOSABEtal}.} \label{fig3}
\end{figure}

\begin{figure}
\resizebox{6cm}{!} {\includegraphics{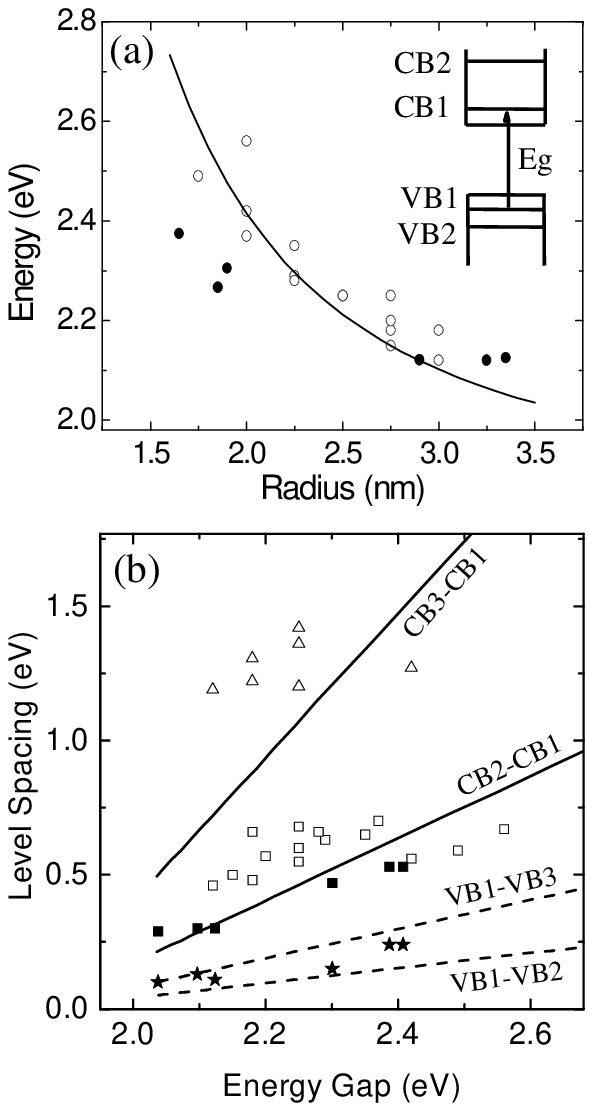}} \caption{(a)
Energy gap versus QR radius. The tunneling data (empty circles)
and the theory (solid line) represent the energy separations
VB1-CB1 (see Figs. 2 and 3). The optical data (solid circles) were
determined from the first absorption peak and corrected for the
electron-hole Coulomb interaction, see text. The inset illustrates
the relevant optical transition from VB1 to CB1. (b) Excited
energy states versus energy gap. The calculated energy level
separations, as denoted, are depicted in the solid and dashed
lines. The respective separations between the tunneling peaks are
denoted by open symbols. Spacing between PLE transitions I and II
and the optical energy gap are presented by solid stars and
squares, respectively. For the optical data, the energy gap was
taken as the detection window corrected for the electron-hole
Coulomb interaction.} \label{fig4}
\end{figure}

The contribution of high $m$ states was not clearly resolved in
our spectra.  In the PLE data, such a contribution possibly
manifests itself in the broadening of the peaks, as compared with
QDs.  As for the tunneling spectra, due to the small wavector
component parallel to the surface associated with the tunneling
electron, coupling to higher $m$ values may be suppressed. We
note, however, that our spectra often exhibit a broad background
on top of which the peaks are superimposed (e.g., the middle
curves in Fig. 2).

In conclusion, optical and tunneling spectroscopic data for CdSe
QRs show that the level structure is dominated by radius rather
than by the length.  This behavior, manifested also in the
theoretical calculations, reveals the quasi one-dimensional nature
of the QR even for aspect ratios as small as three.  This allows
one to select a desired QR length and tune its optical or
electrical properties at will, using the diameter.  This ability
is of significant importance for future nanotechnology
applications of QRs \cite{BaninAdvMatEtal, AlivisatosSollarCells}.

 \acknowledgments
{This work was supported in parts by the US-Israel Binational
Foundation, the DIP foundation and by the Israel Science
Foundation. }


\end{document}